\documentclass[conference]{IEEEtran}
\IEEEoverridecommandlockouts
\usepackage{cite}
\usepackage{subcaption}
\usepackage{amsmath,amssymb,amsfonts}
\usepackage{algorithmic}
\usepackage{graphicx}
\usepackage{textcomp}
\usepackage{xcolor}
\usepackage{todonotes}
\usepackage[breaklinks=true]{hyperref}
\usepackage{breakurl}
\usepackage{booktabs}
\usepackage{listings}
\usepackage{xcolor}
\usepackage{mathtools}
\usepackage{fancyhdr}

\definecolor{eclipseStrings}{RGB}{42,0.0,255}
\definecolor{eclipseKeywords}{RGB}{127,0,85}
\colorlet{numb}{magenta!60!black}

\lstdefinelanguage{json}{
    basicstyle=\footnotesize\ttfamily,
    commentstyle=\color{eclipseStrings}, 
    stringstyle=\color{eclipseKeywords}, 
    numbersep=8pt,
    showstringspaces=false,
    breaklines=true,
    frame=lines,
    string=[s]{"}{"},
    comment=[l]{:\ "},
    morecomment=[l]{:"},
    literate=
        *{0}{{{\color{numb}0}}}{1}
         {1}{{{\color{numb}1}}}{1}
         {2}{{{\color{numb}2}}}{1}
         {3}{{{\color{numb}3}}}{1}
         {4}{{{\color{numb}4}}}{1}
         {5}{{{\color{numb}5}}}{1}
         {6}{{{\color{numb}6}}}{1}
         {7}{{{\color{numb}7}}}{1}
         {8}{{{\color{numb}8}}}{1}
         {9}{{{\color{numb}9}}}{1}
}
\def\BibTeX{{\rm B\kern-.05em{\sc i\kern-.025em b}\kern-.08em
    T\kern-.1667em\lower.7ex\hbox{E}\kern-.125emX}}

\begin{document}

\title{A Toolbox for Design of Experiments for Energy Systems in Co-Simulation and Hardware Tests\\
\thanks{This work has been supported by the ERIGrid 2.0 project of the H2020 Programme under Grant Agreement No. 870620. The contribution of K. Heussen has also been supported by National Funds through the Portuguese funding agency, FCT - Fundação para a Ciência e a Tecnologia, within project LA/P/0063/2020 (DOI 10.54499/LA/P/0063/2020) under INESC TEC International Visiting Researcher Programme.
\\
\copyright2024 IEEE.~https://doi.org/10.1109/OSMSES62085.2024.10668967
Personal use of this material is permitted. Permission from IEEE must be obtained for all other uses, in any current or future media, including reprinting/republishing this material for advertising or promotional purposes, creating new collective works, for resale or redistribution to servers or lists, or reuse of any copyrighted component of this work in other works.}
}
\author{\IEEEauthorblockN{Jan Sören Schwarz}
\IEEEauthorblockA{\textit{R\&D Division Energy} \\
\textit{OFFIS Institute}\\
Oldenburg, Germany \\
0000-0003-0261-4412}
\and
\IEEEauthorblockN{Leonard Enrique Ramos Perez}
\IEEEauthorblockA{\textit{European Distributed Energy} \\
\textit{Resources Laboratories (DERlab) e.V.}\\
Kassel, Germany \\
0000-0001-7912-453X}
\and
\IEEEauthorblockN{Minh Cong Pham}
\IEEEauthorblockA{\textit{CEA-Liten, INES} \\
Le Bourget du Lac\\
France\\
minh-cong.pham@cea.fr}
\and
\IEEEauthorblockN{Kai Heussen}
\IEEEauthorblockA{\textit{Department of Wind and Energy Systems} \\
\textit{Technical University of Denmark}\\
Kgs. Lyngby, Denmark \\
0000-0003-3623-1372}
\and
\IEEEauthorblockN{Quoc Tuan Tran}
\IEEEauthorblockA{\textit{CEA-Liten, INES} \\
Le Bourget du Lac\\
France\\
quoctuan.tran@cea.fr}
}

\maketitle

\begin{abstract}
In context of highly complex energy system experiments, sensitivity analysis is gaining more and more importance to investigate the effects changing parameterization has on the outcome.
Thus, it is crucial how to design an experiment to efficiently use the available resources.
This paper describes the functionality of a toolbox designed to support the users in design of experiment for (co-)simulation and hardware tests. 
It provides a structure for object-oriented description of the parameterization and variations and performs sample generation based on this to provide a complete parameterization for the recommended experiment runs. 
After execution of the runs, it can also be used for analysis of the results to calculate and visualize the effects. 
The paper also presents two application cases using the toolbox which show how it can be implemented in sensitivity analysis studies with the co-simulation framework ``mosaik'' and a hybrid energy storage experiment.
\end{abstract}



\section{Introduction}

As part of the energy transition towards renewable energy resources, the energy systems are undergoing significant changes. To better understand the implications of such transition both simulations and laboratory tests are performed. In general, but especially in context of high complexity, modeling and simulation and laboratory tests of a system always introduce different kind of uncertainty.
One crucial method for managing this uncertainty is Sensitivity Analysis (SA), which examines how changes in input variables affect the output of a simulation \cite{Zhang2020}. 
The first essential step in this process is identifying sources of uncertainty and expressing them mathematically. 
The most common types of representation of uncertainty are intervals and probability distributions, but also more complex approaches are available \cite{jra_1_2}. 
Based on the description of input uncertainty, a screening of the input factors can efficiently reveal those that significantly affect model or system responses. 
This initial analysis helps to save time and resources by identifying non-influential factors and reducing their impact on response variability in subsequent analyses.
\par
In terms of technical testing, the objective in planning and designing tests is to extract the maximum amount of information given a limited budget. 
In pursuit of this objective, the theory of Design of Experiments (DoE) offers sampling strategies that optimize the information that can be obtained from a restricted number of experiments \cite{fisher1935design,Giunta2003,Kleijnen2015}.
It involves utilizing a variety of statistical tools and concepts to manage variance and fluctuations in test inputs, setups, and results. 
Originally developed for physical experiments, the DoE approach has since expanded to encompass various applications.
The effectiveness of DoE in power systems and Cyber-Physical Energy Systems (CPES) testing has been demonstrated in the ERIGrid project, and a connection to the Holistic Test Description (HTD) method has been established \cite{8405401}. 
The HTD assists in defining the scope of a test and includes also information about the uncertainty \cite{jra_1_2}.
\begin{figure*}[th]
	\centering
	\includegraphics[width=0.9\linewidth]{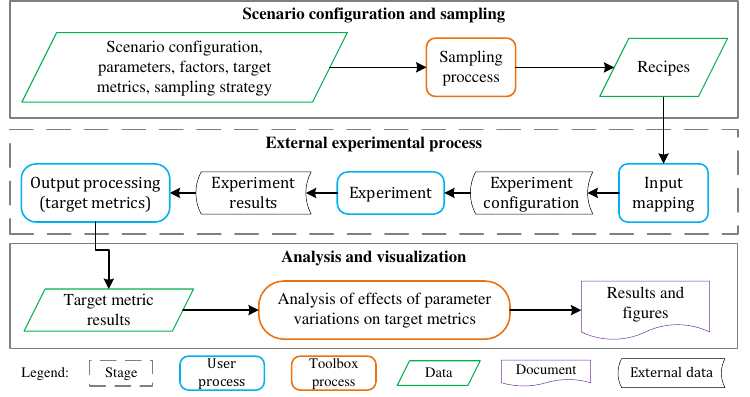}
	\caption{\centering{Sub-processes for the implementation of the developed toolbox within experiments.}}
	\label{fig:methodology}
 \end{figure*}
\par
For the implementation of SA and DoE some libraries exists, like the well-known Sensitivity Analysis Library in Python  (SALib)\footnote{\href{https://salib.readthedocs.io/en/latest/index.html}{https://salib.readthedocs.io/en/latest/index.html}} \cite{Herman2017,Iwanaga2022}, which provides many sampling and analysis methods.
In this work, we present a toolbox for SA and DoE, which uses material from ERIGrid 1.0 \cite{tue2019} and 2.0 \cite{jra_1_2} project, SALib, and additional methods as base to support user in applying SA in simulation but also laboratory tests of energy systems.
It provides a object-oriented data interface for defining parameterization and variability in the experiments inputs and allows to apply different DoE standard methods to provide all needed inputs for a simulation or hardware experiment in one file.
Based on the experiment results and the chosen sampling method, the effects can be analysed and visualized.
\par
The proposed methodology for this toolbox will be described in Section \ref{sec:methodology}.
In Section \ref{sec:usecases} two application cases are shown applying the toolbox to a co-simulation scenario of a multi-energy scenario and a hybrid storage case.
Finally, we conclude the paper in Section \ref{sec:conclusion}.

\section{Methodology}\label{sec:methodology}

The toolbox described in this paper is designed to support the implementation of DoE for both simulation and laboratory experiments of CPES. 
The documentation and implementation are available in an open-access repository at GitHub\footnote{\href{https://github.com/ERIGrid2/toolbox\_doe\_sa}{https://github.com/ERIGrid2/toolbox\_doe\_sa}}. 
The functionality of the developed toolbox can be split into three stages as shown in Figure~\ref{fig:methodology} and described in the following. 
Firstly, the data for the experiment (incl. inputs and parameters) is prepared for the scenarios to be run. 
Secondly, the experiment is executed in a simulation software or laboratory experiment. 
Thirdly, the results are analysed and visualised.

\begin{table*}[]
\centering
\begin{tabular}{@{}llll@{}}
\toprule
DoE type & Sampling Method & Analysis Method & Use Case \\ \midrule
extreme\_points       &   Minimum and maximum values of interval  &  ANOVA & Screening \\
sobol       &   Sobol sequence   &  Meta model & Investigate impact of factors in more detail \\
LHS       &  LHS   &   Meta model & Investigate impact of factors in more detail \\
OAT       &   For each factor one scenario with min and max value   &  OAT & Initial screening of factors \\
sobol\_indices       &    Sobol indices sampling   &  Sobol indices & More detailed and time consuming screening \\
fast       &   eFAST sampling    &  FAST & More detailed and time consuming screening \\
distribution\_and\_discrete       &   Random based on distribution or discrete set &  Not yet implemented     &    -     \\ \bottomrule
\end{tabular}
\caption{Configuration of sampling and analysis methods of toolbox.}
\label{tab:analysis_methods}
\end{table*}

\subsection{Scenario configuration and sampling}
\begin{figure}[h]
	\centering
\lstset{basicstyle = \small}
\begin{lstlisting}[language=json]
{   "samples": 64,
    "doe_type": "sobol",
    "basic_conf": {
        "scenario_name": "test_1",
        "folder_temp_files": "output\\temp_files",
        "end": 604800,
        "step_size": 60,
    },
    "entities_parameters":{
        "storage_tank":{
            "INNER_HEIGHT": 7.9,
            "INSULATION_THICKNESS": 0.1,
        },
    },
    "variations_dict": {
        "storage_tank":{
            "INNER_DIAMETER": [1,8]
        },
    },
    "target_metrics": [
        "electricity_balance_mwh",
    ]
}
\end{lstlisting}
	\caption{Example of a toolbox scenario configuration file.}
	\label{fig:configurationFile}
 \end{figure}

As first step, the user has to define the scenario configuration in a JSON file, which contains five parts as shown in Figure \ref{fig:configurationFile} with an example from the first application case described in Section \ref{sec:usecases_MEB}. More detailed information can be found in the documentation in the repository or in \cite{jra_1_2}.

At the root level, the type of DoE and sample size are defined. 
The DoE type can be specified as shown in Table~\ref{tab:analysis_methods} and contains a sampling strategy and the corresponding analysis method.
The following sampling approaches are available: 

\begin{itemize}
    \item \textbf{Sobol sequences} (quasi-random sequences) which allow the sampling result to be reproducible. Thus, it is also extendable if later a larger number of samples is needed.
    \item \textbf{Latin Hypercube} which is a space filling sampling approach which divides the space in sections with same probability and reduces the clumps, that might occur with completely random sampling.
    \item \textbf{Extreme point sampling} which only uses the minimum and maximum values of the defined intervals and combines them.
    \item \textbf{One-at-a-time (OAT)} which is a SA approach in which only one parameter is changed per simulation run. Thus, it only needs a relative small number of simulation runs, but the results should not be overrated as discussed in \cite{SALTELLI20101508}.
    \item \textbf{Sobol Indices} which are used for global variance-based SA, to find out which parameter have the most effect on the simulation results \cite{Saltelli2010}. To achieve meaningful results, the sample size should be at least 1000, but preferably higher. Depending on the number of factors, the number of scenarios to be executed can get very high.
    \item \textbf{The extended Fourier Amplitude Sensitivity Test (eFAST)} is another global variance-based SA method. The results are comparable to the results with Sobol indices method, but eFAST is slightly more efficient \cite{eFAST}.
    \item Sampling randomly from a \textbf{discrete} set of values or a probability \textbf{distribution}.\\
\end{itemize}
General information about the scenario is defined in the \texttt{basic\_conf} part which contains, for example, a name for the scenario and a path to a folder for temporary files. 
Also specific information for the experiment setup are defined, such as the simulation time and step size of simulators for the (co-)simulation.
\par
Standard parameterization for the experiment simulation models or devices is described in the \texttt{entities\_parameters} part. The content is structured by objects (e.g., simulators or devices) and their parameters with the default value.
\par
In the \texttt{variations\_dict} part, the variation ranges of the factors are specified, i.e., the parameters to be varied to assess the sensitivity will be defined at this point.
Usually a range is defined, but for some methods also distributions or discrete sets can be used here.
Guidelines on how to find and define variations can be found in \cite{jra_1_2}
\par
Finally, the target metrics which are under analysis have to be defined as a list in the \texttt{target\_metrics} part.
\par
Based on the configuration file the \texttt{sample generation} is executed and the so-called recipes are written to a file. 
These recipes define all parameter values for each specific experiment run, combining the default values with the generated samples, and should contain all information needed to run the experiments. 

\subsection{External Experimental Process}

Once the set of recipes was generated, this data has to be adapted to the input structure of the software utilised for performing the experiment. 
Due to the object-oriented approach, the parameter values can be directly mapped to the subsystems of the experiment, independently whether these are different simulation components in a co-simulation or different hardware devices in a laboratory.
At this point, all scenarios previously generated are simulated within the model under study. 
As example in the repository the co-simulation framework mosaik was used \cite{Ofenloch2022}, but any other simulation tool or a experimental setup in a laboratory could be used alternatively.
During the experiment, results are produced which have to be brought in the right format to make them available for the analysis.
Thus, all defined target metrics have to be calculated and provided as JSON, CSV, or HDF5 file for the third stage.
The results file for the analysis has to contain a list of data sets for the different experiment runs which were executed.
For each run the values of the parameters, which were used to start the experiment, and the calculated target metrics have to be provided.

\subsection{Analysis and Visualization}

Finally, in the third stage the previously prepared results from the experiment runs are analysed.
Depending on the chosen sampling method an adequate analysis method is used (see Table \ref{tab:analysis_methods}) and results printed out or stored as CSV tables or images.

\begin{itemize}
    \item \textbf{Analysis of variances (ANOVA)}: This is a method which examines whether the variation in one (independent) factor has a visible effect in the dependent variable by analysing the variations in the factor of different group samples \cite{bib:wapole}. 
    If multiple target metrics have to be considered, ANOVA is executed several times the different combinations of factors. But this might miss patterns which exists between multiple target metrics. Thus, an additional Multivariate ANOVA (MANOVA) is executed for these cases.
    The important outcome of the ANOVA is the p-value, which has to be lower than a specific threshold value (usually 0.05) to indicate a significant effect, as can be seen in the example in Table~\ref{tab:anova}.
    \item \textbf{Meta model}: The goal of a meta model, is to find a simplified model describing the behaviour of the initial model. In the toolbox Kriging method is used based on Gaussian Processes to solve regression and probabilistic classification by calculating the mean and standard deviation from samples~\cite{scikit2024}.
    The meta model is used for a 3D visualization of two factors and one target metric (see example in Figure~\ref{fig:meta_model}).
    \item \textbf{OAT}: The effects of changing one factor at a time are analysed and visualized. The results should not be overrated, as this is a very basic approach and might miss effects between different factors and target metrics \cite{SALTELLI20101508}.
    \item \textbf{Sobol Indices}: Variance-based method for global SA in which the system is analyzed from a probabilistic perspective, where the input model is characterized by a joint probability density function. This relation is expressed as the so call Sobol indexes which represents the contribution of the interaction between all the variables in the system~\cite{Tosin2020}. As results the first-order sensitivity coefficient (S1) and total effect index (ST) are visualized (see example in Figure~\ref{fig:Sobol_all}).
    \item \textbf{eFAST}: In the eFAST each parameter of the model is singled out by assigning a characteristic frequency through a search function. So, the variance contribution to the input model of each one can be analysed separately by this frequency~\cite{Xu2008}.
\end{itemize}

\section{Application Cases}\label{sec:usecases}

In this section, two scenarios demonstrate the application of the toolbox: 
The first is a multi-energy co-simulation scenario implemented in the co-simulation framework mosaik. 
The second is a hybrid storage simulation case, which is based on a hardware experiment and shows the generality of the toolkit.
\subsection{Multi-energy benchmark}\label{sec:usecases_MEB}

As first demonstrating application case a multi-energy benchmark is used as example for implementing a SA with the toolbox. 
The whole application case was described already in detail in \cite{Schwarz2013} and only some parts are shown here as example to visualize results of the toolbox.
It is a co-simulation scenario integrating both electricity and district heating infrastructure creating a multi-energy system. 
It includes a power-to-heat facility acting as a connection point between a low-voltage distribution network and a localized segment of a heating network. This facility serves a dual purpose: utilizing surplus Photovoltaic (PV) generation from the local area to enhance electrical network stability and simultaneously contributing to the thermal network's supply. The chosen scenario involves implementing a local energy community with the goal of using excess local PV generation to operate a power-to-heat facility. Details about this simulation scenario can be found in \cite{DJRA101} and \cite{Widl2022}. 
 
 The objective of the control system is to eliminate the mismatches between energy demand and supply and the voltage spikes at the end of the transmission lines. To address this objective, a voltage control method adjusts the heat pump's consumption target based on voltage monitoring. Meanwhile, a flex heat control scheme determines whether heat should be supplied from the grid or the power-to-heat facility based on tank temperature and heat pump power.
\begin{figure}[t]
    \centering
    \begin{subfigure}{.5\textwidth}
        \centering
        \includegraphics[width=1\linewidth]{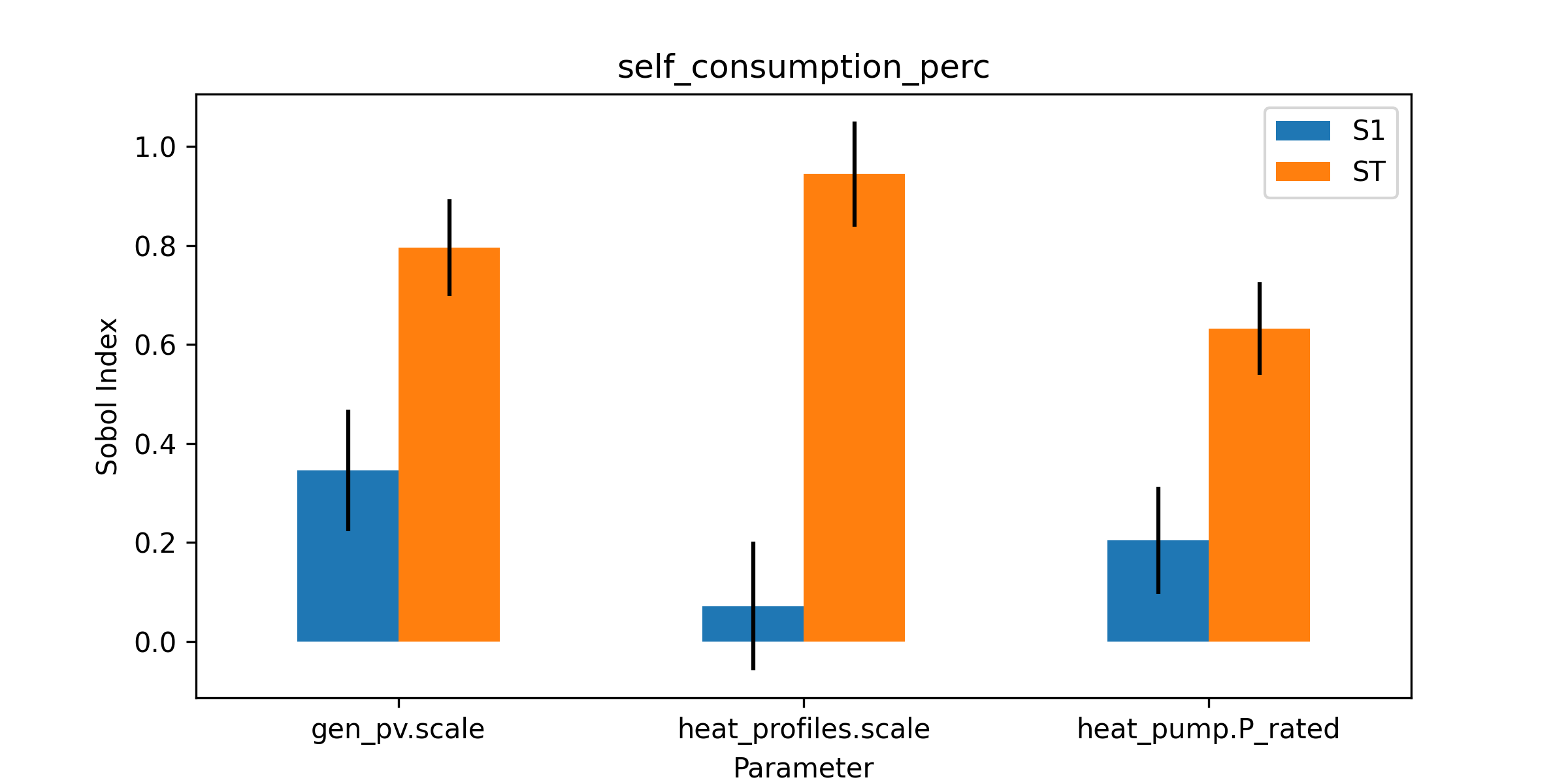}
        \caption{Sobol indices for self-consumption of electricity.}
        \label{fig:Sobol indices for self-consumption of electricity}
    \end{subfigure}
    \newline
    \begin{subfigure}{.5\textwidth}
        \centering
        \includegraphics[width=1\linewidth]{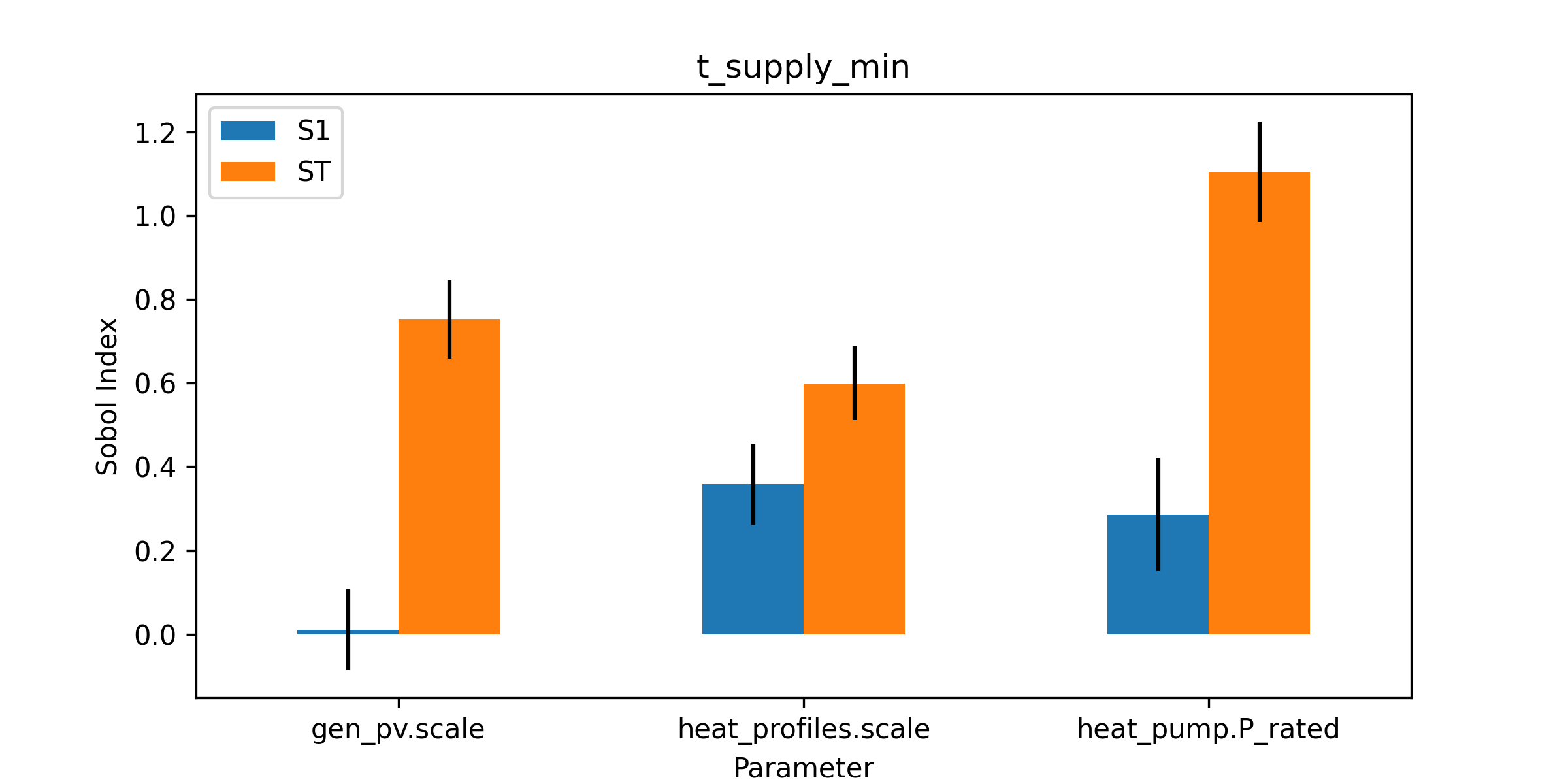}
        \caption{Sobol indices for minimum supply temperature.}
        \label{fig:Sobol indices for critical node temperature of heat}
    \end{subfigure}
    \caption{Sobol indices for different target metrics\protect\cite{Schwarz2013}.}
    \label{fig:Sobol_all}
\end{figure}
\begin{figure}[htbp]
    \centering
    \begin{subfigure}{0.9\linewidth}
        \centering
        \includegraphics[width=1\columnwidth, trim={3.5cm 1cm 2cm 1.9cm}, clip]{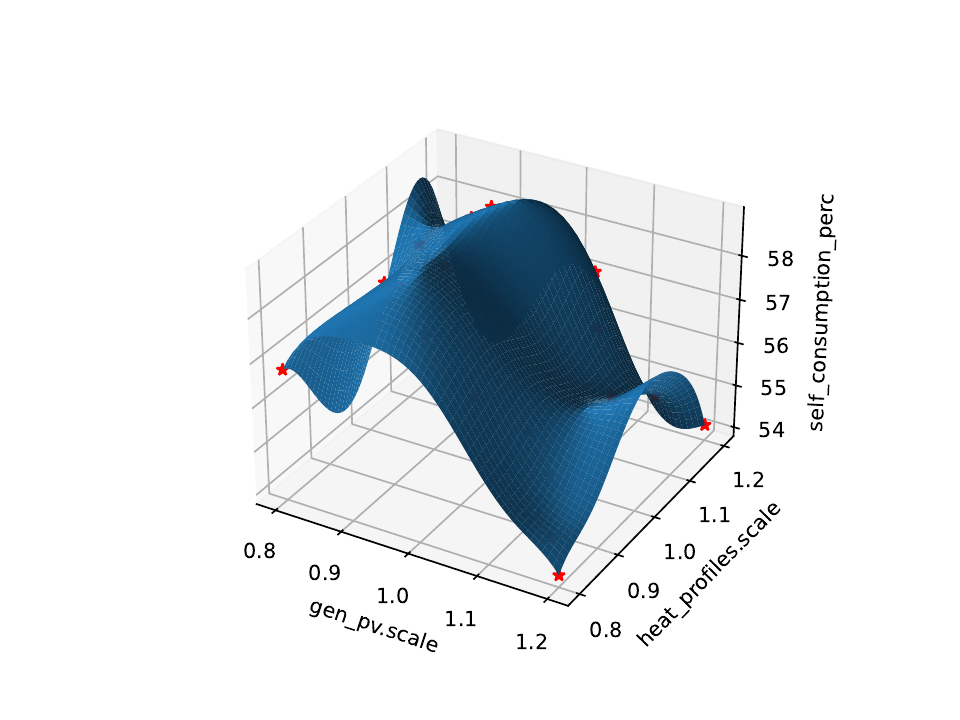}
        \caption{Self-consumption in percent.}
        \label{fig:meb_scaling_pv_selfConsumption}
    \end{subfigure}
    \begin{subfigure}{0.9\linewidth}
        \centering
        \includegraphics[width=1\columnwidth, trim={3.5cm 1cm 2cm 1.9cm}, clip]{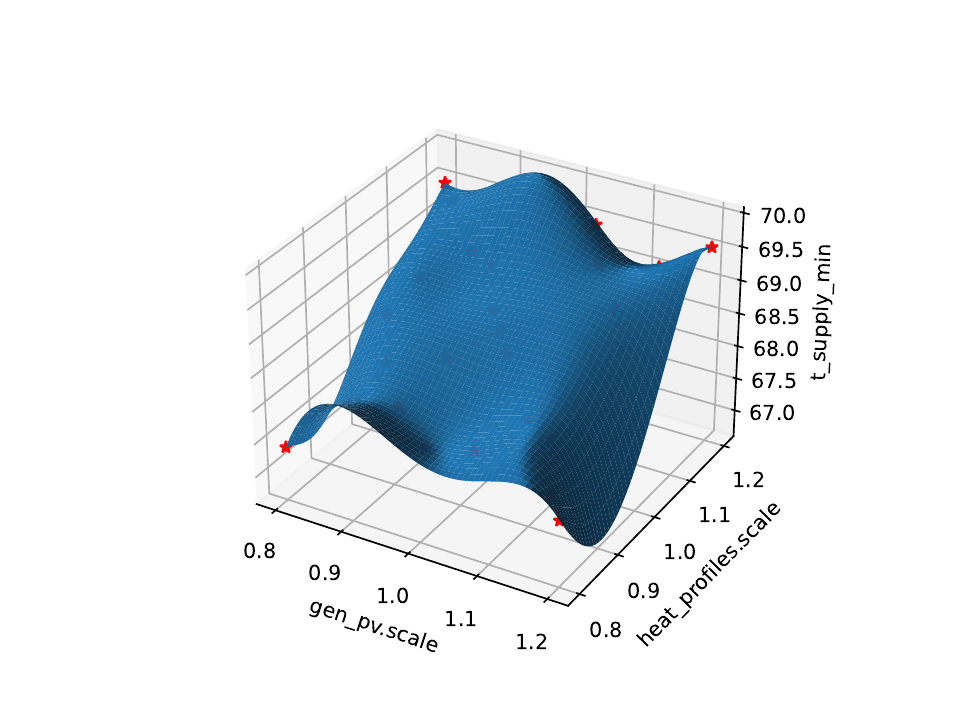}
        \caption{Minimum supply temperature.}
        \label{fig:meb_scaling_pv_Tsupplymin}
    \end{subfigure}
    \caption{Impact of scaling the PV systems and heat demand.}
    \label{fig:meta_model}
\end{figure}

The execution of the multi-energy benchmark needs -- depending on the hardware -- roughly 10 minutes.
Thus, in a first step, a OAT SA was applied.
But this can only be a first indicator for the impact different factors have, because OAT has only limited explanatory power \cite{SALTELLI20101508}.
Thus, the factors for further investigation were chosen based on the OAT results and expert knowledge about the scenario as described in detail in \cite{Schwarz2013}.
Figure \ref{fig:Sobol_all} showcases results from applying the toolbox, demonstrating the Sobol indices Global Sensitivity Analysis (GSA) method. This involves simulating various factor combinations to understand their impact on target metrics. Utilizing 1024 samples, the analysis computes the first-order and total effect indexes (S1 and ST). 
In Figure~\ref{fig:Sobol indices for self-consumption of electricity}, Sobol indices reveal that PV scaling has the greatest first-order impact on electricity self-consumption, followed by the power of the heat pump, while heat profile scaling has a minor first-order effect. 
The configuration file for this simulation can be found also in the repository \footnote{\burl{https://github.com/ERIGrid2/toolbox_doe_sa/blob/main/simulation_configurations/simulation_parameters_inter_domain_scenarioparams.json}}.
In Figure~\ref{fig:Sobol indices for critical node temperature of heat}, Sobol indices show that the heat profile has the greatest first-order impact on the minimal supply temperature, followed by the power of the heat pump, while the PV scaling has a minor first-order effect.

To further investigate the impact of these two factors heat profiles scaling with scaling of the PV systems, also a meta model analysis with sobol sequence sampling was implemented \footnote{\burl{https://github.com/ERIGrid2/toolbox_doe_sa/blob/main/simulation_configurations/simulation_parameters_meta_OSMSES.json}}.
The results are shown in Figure \ref{fig:meta_model}, with the two factors on the x- and y-axis and the target metric on the z-axis.
A meta model analysis visualized in Figure \ref{fig:meb_scaling_pv_selfConsumption} provides further confirmation that the dominant factor influencing electricity self-consumption, as indicated by the previous Sobol indices results, is the scaling of PV. 
Similarly, Figure \ref{fig:meb_scaling_pv_Tsupplymin} shows that the heat profile seems to have a significant impact on the minimum supply temperature target metric, while an effect of PV scaling is not visible.

\subsection{Hybrid storage case}

In the second application case a hybrid energy storage system (HESS) is analysed, which is capable of delivering frequency service and energy arbitrage. 
The focus of this work is to analyse and test the overall hybrid energy storage system to identify significant control parameters and to quantify the trade-offs associated with the relative sizing of three different storage technologies. 

\begin{figure*}[h]
    \centering
    \includegraphics[width=0.85\linewidth]{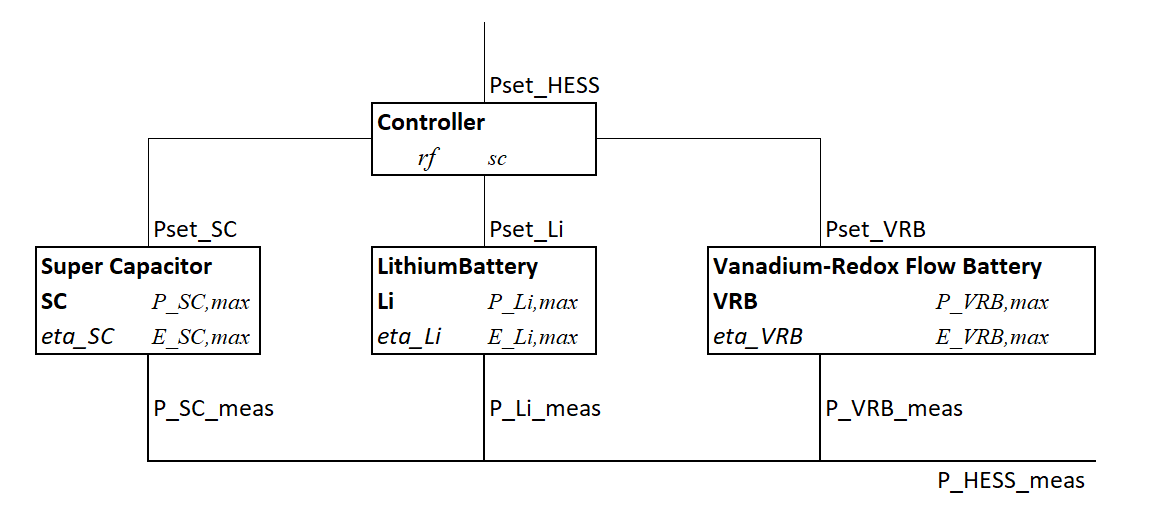}
    \caption{Overview of the HESS}
    \label{fig:HESS}
\end{figure*}

The case is based on the HESS laboratory setup developed in the DK project "Hybrid Energy Storage" 
and consists of three energy storage systems (ESS) and a custom controller. The storage technologies are Super Capacitor (SC), Lithium-Ion battery (Li), and a Vanadium-Redox flow Battery (VRB). 
Matching the laboratory setup, there is a simulation framework that uses the same controller, but simplified ESS models, which will be used for reference in this study. 
An outline of the system structure is presented in Figure~\ref{fig:HESS}.

The main benefits identified for the hybrid system configuration as opposed to the single storage technologies are associated with improved operational efficiency (\textit{Losses\_hess}) and the extension of Li lifetime (\textit{Degradation\_li}), which is calculated as a proxy based on depth-of-discharge.
Thus, the custom controller attempts to a) reduce losses and b) minimize the cycling (energy throughput) of the Li by optimal power  allocation. 
The controller aims to account for the non-constant (and nonlinear) efficiency trade-off between Li and VRB units, while minimising the Li power cycles by offloading power fluctuations to the SC. 
The control strategy is a basic rule-based controller. 
This study illustrates several DoE methods applied to this setup, however, only using the available simulation setup as a black-box, and focusing on the delivery of frequency ancillary services (thus omitting energy arbitrage).

\subsubsection{Parameter analysis and factor selection}

The metrics, \textit{Losses\_hess} and \textit{Degradation\_li} are evaluated for each simulation run. 

The relevant controller parameters are the \textit{restoration factor} ($r\!f$) and the \textit{VRB capacity fraction} ($c\!f$).  
For the ESS, the peak power capacity of all three storage units may be varied, $P^\textrm{max}_{i}~\forall i\in\{\textrm{SC,Li,VRB}\}$%
, is reduced to two relative parameters $a_{SC}$ and $a_{Li}$, by fixing the total HESS system capacity $P^\textrm{max}_{\textrm{HESS}}$ to a constant:
$$P^\textrm{max}_{i}=  P^\textrm{max}_{\textrm{HESS}}(a_{SC}+a_{Li}+(1-a_{SC}-a_{Li}))~\forall i\in\{\textrm{SC,Li,VRB}\} $$

A source of variability is the specific frequency profile used in the simulation. 
As \textbf{nuisance factor}, for each simulation parameter combination, $n_{\textrm{PP}}$ samples can be drawn from a set of 1h frequency profiles. 
Another source of variability is the initial state of charge of each ESS. Very high and low values for $SOC_{\textrm{SC,Li,VRB},init}$ will cause distortion of the results, and are also not realistic, when the storages are managed by an optimal scheduling routine. To avoid excessive sampling, a \textbf{blocking strategy} was implemented which allows to reset the storage level every $n_R$ experiments by a separate procedure.

Two nuisance factors have been identified: the initial state of charge of the respective ESS, and the power profile $P_{req,t}$. Initial experiments have shown that for  parameter $n_R = 2$, and $SOC_{\textrm{SC,Li,VRB},init}=0.5$ the effect of the initial state of charge can be neutralised. The requested power profile $P_{req,t}$ will be  sampled $n_{\textrm{PP}}=5$ times for robustness, as described above.

\subsubsection{Fast analysis of parameter impact with ANOVA method}
The initial investigation attempts to capture which factors have significant impact on the considered target metrics. 
As factors we assess two controller parameters and the relative power capacity of the three storages. 
The \textit{controller} parameters are: Restoration factor $r\!f\in\left[2,4\right]$, VRB capacity fraction $c\!f\in\left[0.3,0.7\right]$; for the ESS the selected parameters and ranges are $P^\textrm{max}_{\textrm{HESS}} = 22.5~kW$ and $a_i\in\left[0.1,0.45\right]$ for $i\in\{\textrm{SC,Li}\}$.  

The ANOVA analysis results are summarised in Table~\ref{tab:anova}. The analysis clearly demonstrates that the parameters $r\!f$ and $c\!f$ are the only ones with significant effects ($p<0.05$). 

As multiple target metrics are considered, additionally, a MANOVA is executed, which confirmed the results of ANOVA: $r\!f$ and $c\!f$ have a significant effect on the target metrics, while $a_{SC}$ and $a_{Li}$ are slightly above the significance threshold. 

\begin{table}
    \centering
    \caption{ANOVA Results for HESS}
\begin{tabular}{lllrr}
\toprule
 & factor & target\_metric & F & p \\
\midrule
0 & $a_{SC}$ & Losses\_hess & 0.000088 & 0.992576 \\
1 & $a_{Li}$ & Losses\_hess & 0.000088 & 0.992576 \\
2 & $r\!f$ & Losses\_hess & 0.007048 & 0.933460 \\
3 & $c\!f$ & Losses\_hess & 59.907739 & 0.000000 \\
4 & $a_{SC}$ & Degradation\_li & 0.084187 & 0.773005 \\
5 & $a_{Li}$ & Degradation\_li & 0.084187 & 0.773005 \\
6 & $r\!f$ & Degradation\_li & 20.544288 & 0.000041 \\
7 & $c\!f$ & Degradation\_li & 2.113200 & 0.152823 \\
\bottomrule
\end{tabular}
\label{tab:anova}
\end{table}

\begin{figure}[t]
    \centering
    \begin{subfigure}{1\linewidth}
        \centering
        \includegraphics[width=1\textwidth]{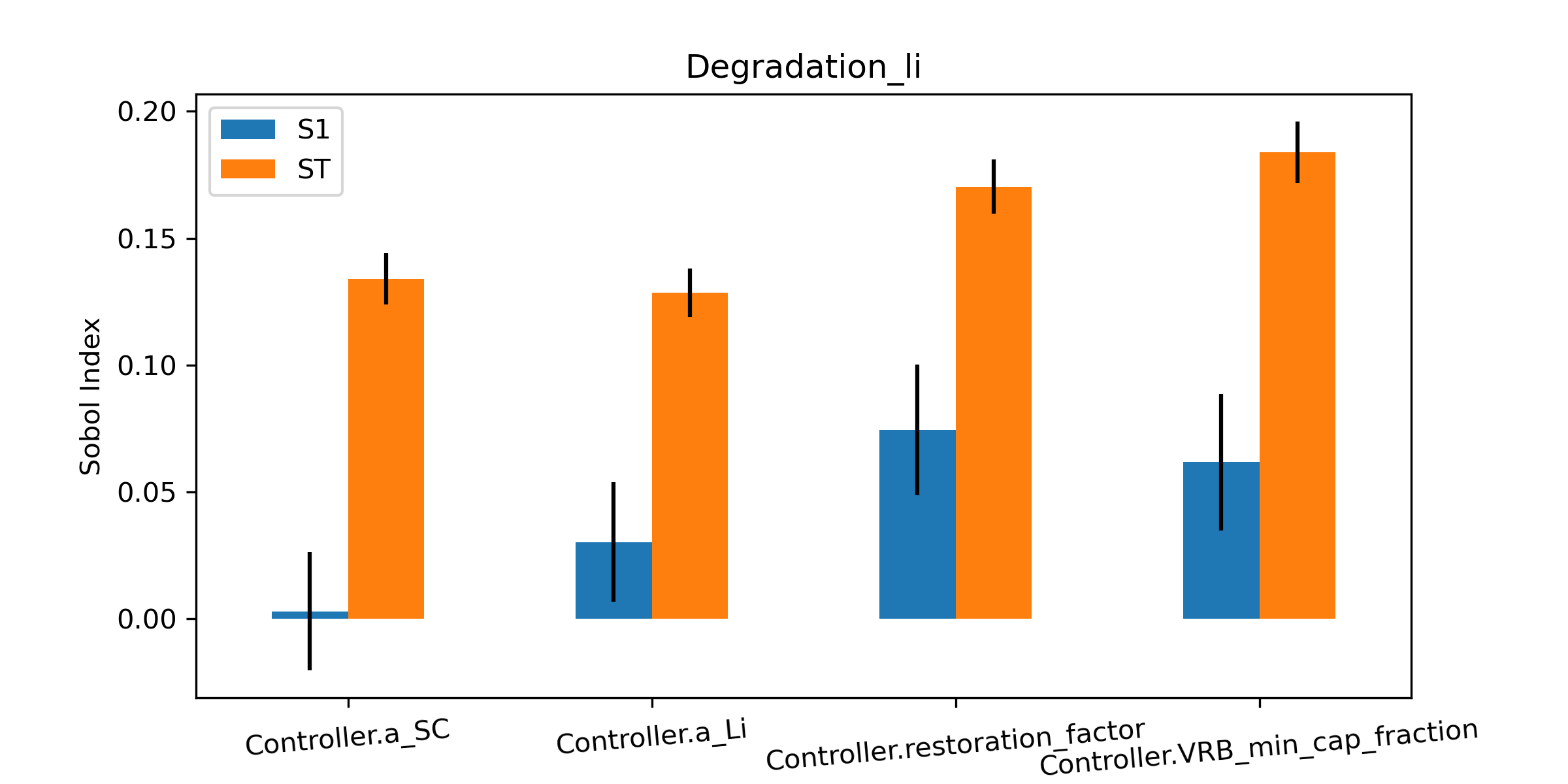}
        \caption{Sobol indices for Lithium Battery Degradation.}
        \label{fig:Sobol_indices_degradation}
    \end{subfigure}
    \newline
    \begin{subfigure}{1\linewidth}
        \centering
        \includegraphics[width=1\textwidth]{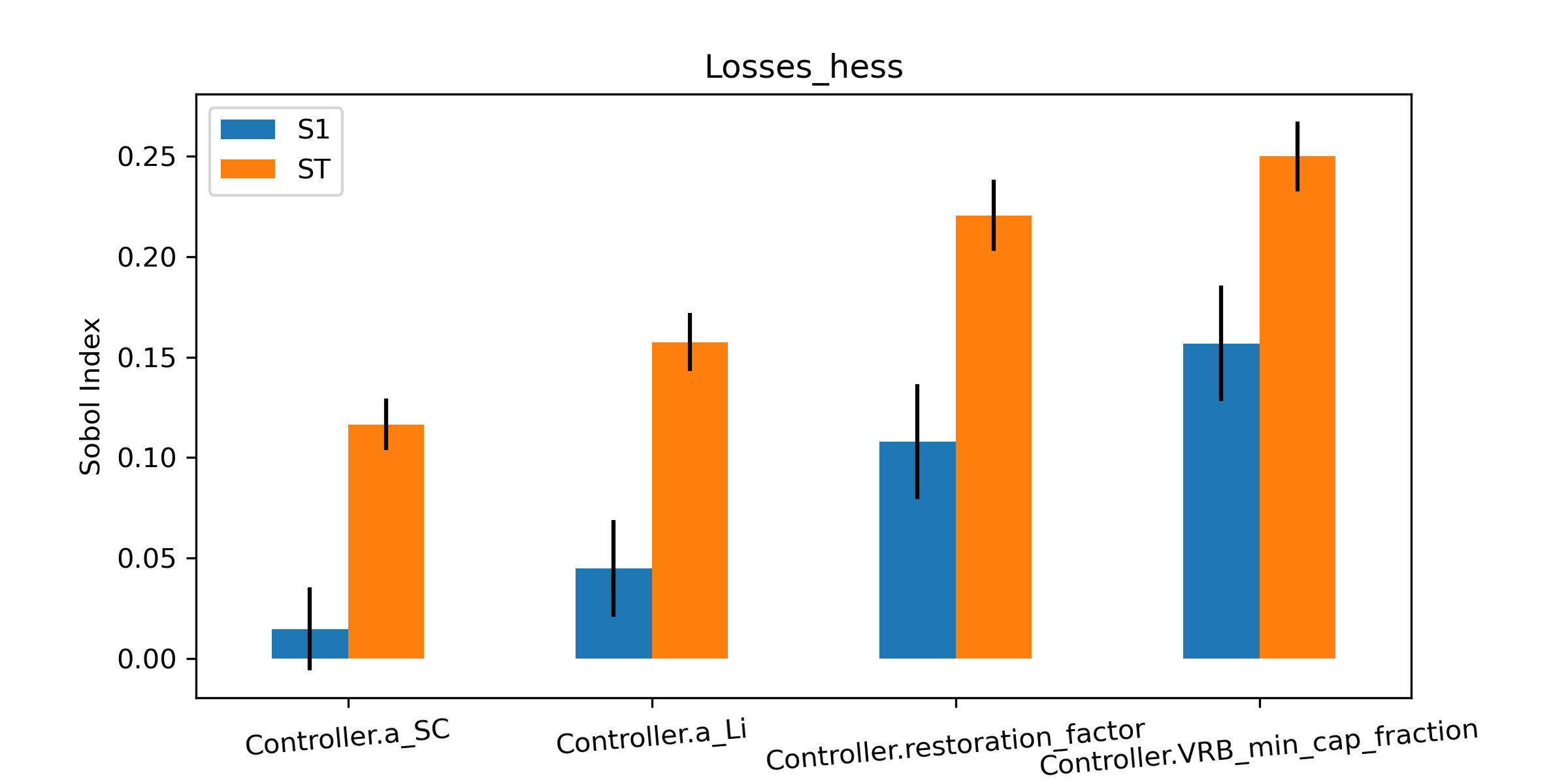}
        \caption{Sobol indices for Losses.}
        \label{fig:Sobol_indices_losses}
    \end{subfigure}
    \caption{Sobol indices for different target metrics of HESS.}
    \label{fig:Sobol_all_HESS}
\end{figure}

\subsubsection{Screening non-linear interactions using Sobol Indices}
To clarify the effects observed in the ANOVA and MANOVA analyses, we further calculate the Sobol indices for all considered parameters. This study should clarify in which form the "insignificant" parameters $a_{SC}$ and $a_{Li}$ contribute to the observed variability. 
The concept of Sobol indices \cite{sobol2001global,Tosin2020} provides a direct means of method for screening the direct and indirect (higher-order interaction) effects of multiple parameters. This study therefore addresses the same parameters and ranges as noted above, however, considering a much larger number of parameter combinations $n=512$, for which the DoE toolbox generated $n_p=3072$ parameter combinations. After conversion to the simulation environment, the addition of the reset blocks ($n_{reset}= 1536$) and multiplying with $n_{\textrm{PP}}=5$, a total of $23040$ simulation runs were executed and evaluated, hereby discarding the invalid "reset" runs. These results were then passed back to the DoE toolbox, generating the results presented in Fig.~\ref{fig:Sobol_all_HESS}. 

The result from this analysis on the first order (blue bars), confirms the observations from the ANOVA study. Hovever, the tall orange bars for most factors indicate a strong interaction effects, indicating possible nonlinear effects that could be studied further through in-depth data analysis. The since the two controller parameters $r\!f$ and $c\!f$ here dominate the results, both on S1 and ST effects, it would be advised to identify the characteristic trade-offs among these parameters first to fix the optimal controller parameters before further studying the sizing effects in detail.

\section{Conclusion}\label{sec:conclusion}
The presented toolbox provides support in setting up DoE for experiments. 
Based on the object-oriented configuration, sample generation is done and can be used for execution of simulation and hardware experiments.
The results can be analyzed and visualized in the toolbox again as demonstrated with two application cases.

The toolbox supports already many different sampling and analysis methods, but it can be further extended in the future to support more. 
For example, specific handling of known and controllable nuisance factor could be integrated, which was done manually in the second application case.
For this, blocking can be used to eliminate the effect of a nuisance factor by building blocks per value of the nuisance factor.
The toolbox has been developed with the described multi-energy benchmark as first exemplary application case.
The generality of the approach is shown with the second application case, but some parts might still have to be implemented more general to be also usable for other types of experiments and adapters to more execution framework would offer even more potential use cases in the future.

For the usage of the toolbox a clear identification and description of uncertainties is required.
It would be beneficial to integrate already available information which might be available already from other sources.
For example, recent extensions of the HTD provide a template for describing uncertainty in an experiment \cite{jra_1_2} and automatic import would facilitate a direct integration in the experiment development processes.


\bibliographystyle{IEEEtran}
\bibliography{literature}

\begin{thebibliography}{10}
\providecommand{\url}[1]{#1}
\csname url@samestyle\endcsname
\providecommand{\newblock}{\relax}
\providecommand{\bibinfo}[2]{#2}
\providecommand{\BIBentrySTDinterwordspacing}{\spaceskip=0pt\relax}
\providecommand{\BIBentryALTinterwordstretchfactor}{4}
\providecommand{\BIBentryALTinterwordspacing}{\spaceskip=\fontdimen2\font plus
\BIBentryALTinterwordstretchfactor\fontdimen3\font minus
  \fontdimen4\font\relax}
\providecommand{\BIBforeignlanguage}[2]{{%
\expandafter\ifx\csname l@#1\endcsname\relax
\typeout{** WARNING: IEEEtran.bst: No hyphenation pattern has been}%
\typeout{** loaded for the language `#1'. Using the pattern for}%
\typeout{** the default language instead.}%
\else
\language=\csname l@#1\endcsname
\fi
#2}}
\providecommand{\BIBdecl}{\relax}
\BIBdecl

\bibitem{Zhang2020}
J.~Zhang, J.~Yin, and R.~Wang, ``{Basic framework and main methods of
  uncertainty quantification},'' \emph{Mathematical Problems in Engineering},
  vol. 2020, 2020.

\bibitem{jra_1_2}
\BIBentryALTinterwordspacing
J.~S. Schwarz, E.~Schulte, K.~Heussen, J.~Nikoletatos, L.~Ramos, and Z.~Feng,
  ``D-{{JRA1}}.2 methods for holistic test reproducibility,'' May 2024.
  [Online]. Available: \url{https://doi.org/10.5281/zenodo.8081442}
\BIBentrySTDinterwordspacing

\bibitem{fisher1935design}
R.~Fisher, \emph{The Design of Experiments}.\hskip 1em plus 0.5em minus
  0.4em\relax Oliver and Boyd, 1935.

\bibitem{Giunta2003}
A.~A. Giunta, S.~F. Wojtkiewicz, and M.~S. Eldred, ``{Overview of modern design
  of experiments methods for computational simulations},'' in \emph{41st
  Aerospace Sciences Meeting and Exhibit}, no. January, 2003.

\bibitem{Kleijnen2015}
J.~P. Kleijnen, \emph{{Design and Analysis of Simulation Experiments
  (International Series in Operations Research {\&} Management Science)}},
  2015.

\bibitem{8405401}
A.~A. van~der Meer, C.~Steinbrink, K.~Heussen, D.~E.~M. Bondy, M.~Z. Degefa,
  F.~P. Andrén, T.~I. Strasser, S.~Lehnhoff, and P.~Palensky, ``Design of
  experiments aided holistic testing of cyber-physical energy systems,'' in
  \emph{2018 Workshop on Modeling and Simulation of Cyber-Physical Energy
  Systems (MSCPES)}, 2018, pp. 1--7.

\bibitem{Herman2017}
J.~Herman and W.~Usher, ``{SALib}: An open-source python library for
  sensitivity analysis,'' \emph{The Journal of Open Source Software}, vol.~2,
  no.~9, Jan. 2017.

\bibitem{Iwanaga2022}
T.~Iwanaga, W.~Usher, and J.~Herman, ``Toward {SALib} 2.0: {Advancing} the
  accessibility and interpretability of global sensitivity analyses,''
  \emph{Socio-Environmental Systems Modelling}, vol.~4, p. 18155, May 2022.

\bibitem{tue2019}
\BIBentryALTinterwordspacing
T.~V. Jensen, ``Erigrid/na4-summerschool-dtu-2018: v2.0,'' May 2019. [Online].
  Available: \url{https://doi.org/10.5281/zenodo.2837928}
\BIBentrySTDinterwordspacing

\bibitem{SALTELLI20101508}
A.~Saltelli and P.~Annoni, ``How to avoid a perfunctory sensitivity analysis,''
  \emph{Environmental Modelling \& Software}, vol.~25, no.~12, pp. 1508--1517,
  2010.

\bibitem{Saltelli2010}
A.~Saltelli, P.~Annoni, I.~Azzini, F.~Campolongo, M.~Ratto, and S.~Tarantola,
  ``{Variance based sensitivity analysis of model output. Design and estimator
  for the total sensitivity index},'' \emph{Computer Physics Communications},
  vol. 181, no.~2, pp. 259--270, Feb. 2010.

\bibitem{eFAST}
A.~Saltelli, S.~Tarantola, and K.~P.-S. Chan, ``A quantitative
  model-independent method for global sensitivity analysis of model output,''
  \emph{Technometrics}, vol.~41, no.~1, pp. 39--56, 1999.

\bibitem{Ofenloch2022}
A.~Ofenloch, J.~S. Schwarz, D.~Tolk, T.~Brandt, R.~Eilers, R.~Ramirez, T.~Raub,
  and S.~Lehnhoff, ``{{MOSAIK}} 3.0 : {{Combining Time Stepped}} and {{Discrete
  Event Simulation}},'' in \emph{1st {{International Workshop}} on "{{Open
  Source Modelling}} and {{Simulation}} of {{Energy Systems}}"}, Aachen, 2022.

\bibitem{bib:wapole}
R.~E. Walpole, R.~H. Myers, S.~L. Myers, and K.~Ye, \emph{The Analysis of
  Variances Method}, 9th~ed.\hskip 1em plus 0.5em minus 0.4em\relax Pearson
  Education, 2012, pp. 414--416.

\bibitem{scikit2024}
F.~Pedregosa, G.~Varoquaux, A.~Gramfort, V.~Michel, B.~Thirion, O.~Grisel,
  M.~Blondel, P.~Prettenhofer, R.~Weiss, V.~Dubourg, J.~Vanderplas, A.~Passos,
  D.~Cournapeau, M.~Brucher, M.~Perrot, and E.~Duchesnay, ``Scikit-learn:
  Machine learning in {P}ython,'' \emph{Journal of Machine Learning Research},
  vol.~12, pp. 2825--2830, 2011.

\bibitem{Tosin2020}
\BIBentryALTinterwordspacing
M.~Tosin, A.~M.~A. Côrtes, and A.~C. Jr, ``A tutorial on sobol’ global
  sensitivity analysis applied to biological models,'' \emph{Networks in
  Systems Biology: Applications for Disease Modeling}, 2020. [Online].
  Available: \url{doi.org/10.1007/978-3-030-51862-2_6}
\BIBentrySTDinterwordspacing

\bibitem{Xu2008}
\BIBentryALTinterwordspacing
C.~Xu and G.~Z. Gertner, ``A general first-order global sensitivity analysis
  method,'' \emph{Reliability Engineering \& System Safety}, vol.~93, no.~7,
  pp. 1060--1071, 2008, bayesian Networks in Dependability. [Online].
  Available:
  \url{https://www.sciencedirect.com/science/article/pii/S0951832007001342}
\BIBentrySTDinterwordspacing

\bibitem{Schwarz2013}
\BIBentryALTinterwordspacing
J.~S. Schwarz, M.~C. Pham, Q.~T. Tran, and K.~Heussen, ``Scaling analysis in a
  multi-energy system,'' in \emph{2023 Asia Meeting on Environment and
  Electrical Engineering (EEE-AM)}, 2023, pp. 01--06. [Online]. Available:
  \url{https://doi.org/10.1109/EEE-AM58328.2023.10395068}
\BIBentrySTDinterwordspacing

\bibitem{DJRA101}
\BIBentryALTinterwordspacing
S.~D'Arco, A.~De~Paola, E.~Widl, V.~S. Rajkumar, J.~Kamsamrong, P.~Raussi,
  G.~Arnold, D.~Thomas, A.~Marinopoulos, C.~W. Wild, K.~Heussen, E.~Rikos,
  T.~T. Hoang, and A.~F. Cortés-Borray, ``D-jra1.1 benchmark scenarios,''
  Tech. Rep., Jan. 2022. [Online]. Available:
  \url{https://doi.org/10.5281/zenodo.4032691}
\BIBentrySTDinterwordspacing

\bibitem{Widl2022}
E.~Widl, C.~Wild, K.~Heussen, E.~Rikos, and T.-T. Hoang, ``Comparison of two
  approaches for modeling the thermal domain of multi-energy networks,'' in
  \emph{{2022 Open Source Modelling and Simulation of Energy Systems
  (OSMSES)}}, 2022, pp. 1--6.

\bibitem{sobol2001global}
I.~M. Sobol, ``Global sensitivity indices for nonlinear mathematical models and
  their monte carlo estimates,'' \emph{Mathematics and computers in
  simulation}, vol.~55, no. 1-3, pp. 271--280, 2001.

\end{thebibliography}
\end{document}